\documentclass{emulateapj}

\received{} \accepted{}

\shorttitle{A globular cluster in M31}
\shortauthors{Ma et al.}

\begin{document}
\slugcomment{PASP, in press}
\title{Determination of fundamental properties of an M31 globular cluster from main-sequence photometry}

\author{
Jun Ma\altaffilmark{1,2}, Zhenyu Wu\altaffilmark{1}, Song Wang,\altaffilmark{1,3} Zhou
Fan\altaffilmark{1}, Xu Zhou\altaffilmark{1}, Jianghua Wu\altaffilmark{1}, Zhaoji
Jiang\altaffilmark{1} and Jiansheng Chen\altaffilmark{1}}

\altaffiltext{1}{National Astronomical Observatories, Chinese
Academy of Sciences, Beijing, 100012, P. R. China;
majun@vega.bac.pku.edu.cn}

\altaffiltext{2}{Key Laboratory of Optical Astronomy, National Astronomical Observatories, Chinese
Academy of Sciences, Beijing, 100012, China}

\altaffiltext{3}{Graduate University, Chinese Academy of Sciences, Beijing, 100039, P. R. China}

\begin{abstract}

M31 globular cluster B379 is the first extragalactic cluster, the age of which was determined by
main-sequence photometry. In the main-sequence photometric method, the age of a cluster is
obtained by fitting its color-magnitude diagram (CMD) with stellar evolutionary models. However,
different stellar evolutionary models use different parameters of stellar evolution, such as range
of stellar masses, different opacities and equations of state, and different recipes, and so on.
So, it is interesting to check whether different stellar evolutionary models can give consistent
results for the same cluster. \citet{brown04a} constrained the age of B379 by comparing its CMD
with isochrones of the 2006 VandenBerg models. Using SSP models of \citet{bru03} (BC03) and its
multi-photometry, \citet{Ma07} independently determined the age of B379, which is in good
agreement with the determination of \citet{brown04a}. The BC03 models are calculated based on the
Padova evolutionary tracks. It is necessary to check whether the age of B379 which, being
determined based on the Padova evolutionary tracks, is in agreement with the determination of
\citet{brown04a}. So, in this paper, we re-determine its age using isochrones of the Padova
stellar evolutionary models. In addition, the metal abundance, the distance modulus, and the
reddening value for B379 are also determined in this paper. The results obtained in this paper are
consistent with the previous determinations, which including the age obtained by \citet{brown04a}.
So, this paper confirms the consistence of the age scale of B379 between the Padova isochrones and
the 2006 VandenBerg isochrones, i.e. the results' comparison between \citet{brown04a} and
\citet{Ma07} is meaningful. The results obtained in this paper are: the metallicity $\rm{[M/H]} =
\log(Z/Z_\odot)=-0.325$, the age $\tau=11.0\pm1.5$ Gyr, the reddening value $E(B-V)=0.08$, and the
distance modulus $(m-M)_{0}=24.44\pm0.10$.
\end{abstract}


\section{Introduction}

Globular clusters (GCs), relics of some of the earliest phases of star and galaxy formation, can
be analyzed to understand how soon after the Big Bang did the various stellar systems form. The
most direct method for determining the age of a star cluster is main-sequence photometry, in which
the isochrone that minimizes the discrepancies between the observed and calculated sequences can
exactly present the estimated cluster age. However, this method has only been applied to the
Galactic GCs and GCs in the satellites of the Milky Way \citep[e.g.,][]{rich01} before
\citet{brown04a} used this method to constrain the age of an M31 GC B379\footnote{In
\citet{brown04a}, SKHB~312 is used. In this paper, B379 is used following the designations from
the Revised Bologna Catalog (RBC) of M31 GCs and candidates (Galleti et al. 2004, 2006, 2007),
which is the main catalog used in studies of M31 GCs.} based on the CMD reaching more than 1.5 mag
below the main-sequence turn-off. The CMD of B379 was constructed from the extremely deep images
with the Advanced Camera for Surveys (ACS) on the {\sl Hubble Space Telescope} ({\sl HST}).

In general, ages of extragalactic star clusters are obtained by comparing integrated photometry
with models of simple stellar populations (SSPs). For examples, \citet{Ma01,Ma02a,Ma02b,Ma02c} and
\citet{jiang03} estimated ages for star clusters in M33 and M31 by comparing the SSP models of
BC96 (Bruzual \& Charlot 1996, unpublished) with their integrated photometric measurements in the
Beijing-Arizona-Taiwan-Connecticut (BATC) photometric system; \citet{degrijs03a} determined ages
and masses of star clusters in the fossil starburst region B of M82 by comparing their observed
cluster spectral energy distributions (SEDs) with the model predictions for an instantaneous burst
of star formation \citep[see also][]{degrijs03b,degrijs03c}. \citet{bik03} and \citet{bastian05}
derived ages, initial masses and extinctions of M51 star cluster candidates by fitting {\sc
Starburst99} SSP models \citep{Leitherer99} to their observed SEDs in six broad-band and two
narrow-band filters from the Wide Field Planetary Camera-2 (WFPC2) onboard the {\sl HST}.
\citet{Ma06a} estimate ages and metallicities for 33 M31 GCs by comparing between BC03 models
\citep{bru03} and their BATC multi-band photometric data. \citet{Ma06b} derived the age and
reddening value of the M31 GC 037-B327 based on photometric and BC03 measurements in a large
number of broad- and intermediate-band from the optical to the near-infrared. \citet{fan06}
determined new ages for 91 M31 GCs from \citet{jiang03} based on improved photometric data and
BC03 models. In particular, \citet{Ma07} derived the age of B379\footnote{In \citet{Ma07}, S312 is
used.} by comparing its photometric data with BC03 models. The age obtained by \citet{Ma07} is
$9.5_{-0.99}^{+1.15}$~Gyr, which is consistent with the determination of $10_{-1}^{+2.5}$~Gyr by
\citet{brown04a} using the main-sequence photometry.

The nearest large GC system outside the Milky Way is that of the Andromeda Galaxy (= M31), which
is located at a distance of 770 kpc \citep{Freedman90}. The first M31 GC resolved into stars was
studied from the ground by \citet{Heasley88}, who only resolved the red giant branch of G1.
Subsequently, some authors such as \citet{Ajhar96}, \citet{fusi96}, \citet{rich96}, \citet{hfr97},
\citet{Jablonka00}, \citet{Williams01a}, and \citet{Williams01b} have used images from the {\sl
HST}/WFPC2 to construct the CMDs of M31 star clusters in order to determine their metallicities,
reddening values, and ages. However, these CMDs are not deep enough to show conspicuous
main-sequence turn-offs.

Since the luminosity of the horizontal branch (HB) in stellar populations older than about 8 Gyr
is expected to be independent of age and only mildly dependent on metallicity, it is widely used a
distance indicator \citep[see][and references therein]{Gallart05}. In addition, HB stars are
fundamental standard candles for Population II systems, and consequently are important tools for
determining ages of GCs from the main-sequence turn-off luminosity \citep{rich96}.

Two of the first studies about the HB for M31 GCs were that of \citet{rich96} and \citet{fusi96}.
They used the observed data from the Wide Field Planetary Camera-2 (WFPC2) and Faint Object Camera
(FOC) onboard the {\sl HST} to make the first tentative detection of HB in G1, B006, B045, B225,
B343, B358, B405, and B468, deriving the apparent magnitude of the HB for these GCs to be in the
range $25.29<V<25.66$. In addition, \citet{fusi96} firstly presented a direct calibration for the
mean absolute magnitude of the HB at the instability strip with varying metallicity for M31 GCs.

B379 was firstly detected by \citet{sh73} (No.19), and confirmed by \citet{sarg77} ($\rm
No.312=SKHB~312$, which was used by \citet{brown04a}; or $=\rm S312$, which was used by
\citet{Ma07}.) and \citet{battis87} ($\rm No.379=B379$, which is used in the RBC (Galleti et al.
2004, 2006, 2007) and this paper.). B379 is located in the halo of M31, at a projected distance of
about $59\arcmin$ ($\rm =13~kpc$) from the galaxy's nucleus. It is a fact that B379 is a common
halo GC, however, it is among the first extragalactic GCs whose age was accurately estimated by
main-sequence photometry \citep{brown04a} based on its CMD from the extremely deep images observed
with the {\sl HST}/ACS.

In this paper, we re-determine the age, metallicity, reddening value and distance modulus for B379
by comparing its CMD constructed by \citet{brown04a} with isochrones of the Padova stellar
evolutionary models. The paper is organized as follows. In \S 3, we describe the results of
photometric data based on the {\sl HST}/ACS observations for B379. In \S 4, we constrain the age,
metallicity, reddening value, and distance modulus for B379. At last, we will give a summery in \S
5.

\section{Recent works of B379}

\citet{brown04a} used the main-sequence photometry to determine the age of B379 based on the CMD
constructed using the extremely deep images from the {\sl HST}/ACS observations. This CMD reached
more than 1.5 mag below the main-sequence turn-off, and firstly allowed a direct age estimate from
the turn-off for an extragalactic cluster. By comparison to isochrones of \citet{vandenberg06},
\citet{brown04a} derived the age of B379 to be $10_{-1}^{+2.5}$~Gyr. \citet{Ma07} determined the
age of B379 by comparing its multi-color photometric data which including the near-ultraviolet
(NUV) from the Nearby Galaxies Survey (NGS) of the {\sl Galaxy Evolution Explorer} ({\sl GALEX})
\citep{rey05,rey06}, broad-band $UBVR$ \citep{battis87,rhh94}, 9 BATC intermediate-band filters
and Two Micron All Sky Survey (2MASS) $JHK_s$, with SSP models of BC03. These photometric data
constitute the SEDs of B379 covering $2267-20000$\AA. The age of B379 determined by \citet{Ma07}
is $9.5_{-0.99}^{+1.15}$~Gyr, which is consistent with the determination of $10_{-1}^{+2.5}$~Gyr
by \citet{brown04a}. However, BC03 models are based on the Padova evolutionary tracks. So, it is
necessary to compare the age scale between the Padova evolutionary tracks and the Victoria-Regina
isochrones used in \citet{brown04a}, and only if these two evolutionary tracks have the consistent
age scale for B379, the results' comparison between \citet{brown04a} and \citet{Ma07} is
meaningful. As an example, \citet{Ma07} drew the isochrones with 10~Gyr and the solar metallicity,
and found that the matching is very good in the main-sequence (MS) and the subgiant branch (SGB)
\citep[see][for details]{Ma07}. However, we should check whether the age of B379 can be estimated
to be $\sim 10$~Gyr based on the Padova evolutionary tracks. This is one of the key contributions
of the present paper.

\section{Database}

The observed data of B379 in this study are from \citet{brown04a}, who constructed the CMD of B379
using the images observed with the ACS observations in the F606W and the F814W filters. Using the
ACS Wide-Field Camera (WFC), \citet{brown03} obtained deep optical images of a field, $51'$ from
the nucleus on the southeast minor axis of the M31 halo including B379, which are 39.1 hr in the
F606W filter and 45.4 hr in the F814W filter. \citet{brown04a} presented the CMD of B379 based on
these ACS observations. The resulting CMD reached $m_{V}\approx 30.5$ mag, which is the first CMD
of extragalactic clusters reaching more than 1.5 mag below the main-sequence turn-off. These
observations firstly allow a direct age estimate from the turn-off for an extragalactic cluster.
By comparison to isochrones of \citet{vandenberg06}, \citet{brown04a} derived the age of B379 to
be $10_{-1}^{+2.5}$~Gyr. In \citet{brown04a}, the CMD of B379 was constructed from stars within an
annulus chosen to maximize the signal-to-noise and minimize field contamination. Because B379 was
near the field edge and the observations were dithered, the exposure time was not uniform across
the annulus. So, \citet{brown04a} discarded the fraction of annulus ($<0.5\%$) that had half of
the total exposure time but kept the fraction ($<14\%$) that was exposed for $75\%$. In addition,
\citet{brown04a} used extensive artificial star tests to determine the photometric scatter and
completeness as a function of color, luminosity and field position. At last, 1720 stars within the
annulus spanning $100-300$ pixels were retained to produce a much cleaner CMD \citep[see][for
details]{brown04a}. In this paper, we also take these 1720 stars as the member stars of B379 as
\citet{brown04a} did (The data were kindly provided by Dr. Brown).

\section{The age, metallicity, reddening value and distance modulus of B379}

\subsection{Isochrones of stellar evolutionary models}

More than 50 years ago, \citet{sandage53} presented the CMD for the Galactic GC M3 and applied an
evolutionary theory to M3 CMD to give a time interval of $5\times 10^9$ years since the formation
of the main-sequence. From then on, main-sequence photometry is thought the most direct method for
determining ages of star clusters, because the turn-off of the CMD is mostly affected by age
\citep[see][and references therein]{puzia02b}. Stellar evolutionary models from the Padova group
\citep[][and references therein]{bertelli94,Girardi00,Girardi02} and the Victoria-Regina
\citep[][and references therein]{vandenberg00,vandenberg06} are widely used. In the Padova stellar
evolutionary models, \citet{Girardi02} provided tables of theoretical isochrones in such
photometric systems as ABmag, STmag, VEGAmag, and a standard star system, and derived tables of
bolometric corrections for Johnson-Cousins-Glass, {\sl HST}/WFPC2, {\sl HST}/NICMOS, Washington,
and ESO Imaging Survey systems. The complete data-base \citep{Girardi02} covers a very large range
of stellar masses (typically from 0.6 to $120~M_{\odot}$). As a supplement, \citet{Girardi08}
presented several theoretical isochrones including {\sl HST}/ACS WFC. These models
\citep{Girardi02,Girardi08} are computed with updated opacities and equations of state, and
moderate amount of convective overshoot. However, the isochrones are presented for only 6 initial
chemical compositions: ${\rm [Fe/H]}=-2.2490$, $-1.6464$, $-0.6392$, $-0.3300$, $+0.0932$ (solar
metallicity), and $+0.5595$, which are evidently not dense enough. It is fortunate that
\citet{Marigo08} provide tables for any intermediate value of age and metallicity via an
interactive web interface (http://stev.oapd.inaf.it/cmd). We will discuss this web in detail in \S
4.2. The novel feature of the Victoria-Regina models \citep[][and references
therein]{vandenberg00,vandenberg06} is that they provide a wide range of metallicities, i.e.
\citet{vandenberg06} presented seventy-two grids of stellar evolutionary tracks for 32 [Fe/H]
values from $-2.31$ to $0.49$, which are dense enough for studying properties of stellar
populations with different metallicities. In addition, in these models, convective core
overshooting has been treated using a parameterized form of the Roxbergh criterion
\citep{roxburgh78,roxburgh89}, in which the free parameter, $F_{\rm over}$ ($F_{\rm over}$ must be
calibrated using observations.), is assumed to be a function of both mass and metal abundance.

\subsection{Isochrone fitting}

To determine the main characteristics (age and metallicity) of the population in B379, we fit
isochrones to the cluster CMD. We used the Padova theoretical isochrones in the {\sl HST}/ACS WFC
STmag system \citep{Marigo08}. Via an interactive web interface at http://stev.oapd.inaf.it/cmd,
we can construct a grid of isochrones for different values of age and metallicity, photometric
system, and dust properties. We use the default models that involve scaled solar abundance ratios
(i.e., $[\rm{\alpha/Fe]}=0.0$). In performing, the Salpeter initial mass function (IMF)
\citep{salp55} is adopted to match the selection of \citet{Ma07}, who used the high-resolution SSP
models of BC03 computed using the \citet{salp55} IMF; and circumstellar dust is not included. As
we pointed out previously, that by comparison to isochrones of \citet{vandenberg06},
\citet{brown04a} derived the age of B379 to be $10_{-1}^{+2.5}$~Gyr. In addition, the metallicity
of B379 is available: \citet{hbk91} derived $\rm [Fe/H]=-0.7\pm 0.35$ using the strengths of six
absorption features in the cluster integrated spectra; \citet{hfr97} used the {\sl HST}/WFPC2
photometry to construct the deep CMD for B379, and the shape of the red giant branch (RGB) gave an
iron abundance of $\rm [Fe/H]=-0.53\pm 0.03$. These two metallicities obtained from different
methods are consistent. Based on the age and metallicity of B379 obtained by the previous authors
\citep{brown04a,hbk91,hfr97}, we used the interactive web (http://stev.oapd.inaf.it/cmd) to
construct a fine grid of isochrones about ages and metallicities, sampling an age range $8.0 \leq
\tau \leq 13.5$~Gyr at intervals of 0.5 Gyr, and a metal abundance range $0.00250 \leq Z \leq
0.00950$ at intervals of 0.00025 dex. The total metallicity $\rm{[M/H]} = \log(Z/Z_\odot)$ where
$Z_\odot \approx 0.019$, so this abundance range corresponds to $-0.88 \le \rm {[M/H]} \le -0.30$.

We followed the method of \citet{MB07} of finding the best fitting isochrone, i.e. we did this by
locating by eye three fiducial points on the CMD of the cluster: the magnitude and color of the
turn-off, the magnitude of the tight clump of red HB stars, and the color of the RGB at a level
3.0 mag brighter than the level of the turn-off. This latter point was selected simply as a point
lying on the lower RGB at a level intermediate between that of the red end of the SGB and that of
the tight clump of red HB. We then calculated the difference in magnitude between the level of the
turn-off and the level of the tight clump of red HB ($\Delta m_{\rm F814W}$), and the difference
in color between the turn-off and the RGB fiducial point ($\Delta c_{m_{\rm F606W}-m_{\rm
F814W}}$). As \citet{MB07} pointed out that, the difference in magnitude between the level of the
turn-off and the level of the tight clump of red HB is strongly sensitive to cluster age (and
weakly to cluster metallicity), while the difference in color between the turn-off and the RGB
fiducial point is sensitive to both cluster age and metallicity. We determined $\Delta m_{\rm
F814W}=3.77 \pm 0.1$ and $\Delta c_{m_{\rm F606W}-m_{\rm F814W}}=0.28\pm 0.01$.

Second, we calculated the same intervals for all isochrones on the grid, and selected only those
with values lying within certain tolerances of the cluster measurements. In this paper, we adopted
$\pm 0.2$ mag for $\Delta m_{\rm F814W}$ and $\pm 0.02$ for $\Delta c_{m_{\rm F606W}-m_{\rm
F814W}}$. We fit the selected isochrones to the CMD by eye. At the same time, we calculated the
offsets in magnitude and color required to align the turn-off of the isochrone with that of the
CMD, and the offsets in magnitude required to align the tight clump of red HB of the isochrone
with that of the CMD, and the offsets in color required to align the RGB fiducial point of the
isochrone with that of the CMD. We then averaged the offsets in magnitude and in color and applied
to overplot the isochrone on the CMD, and identified the best fitting isochrone by eye. The
resulting offsets $\delta m_{\rm F814W}$ and $\delta c_{m_{\rm F606W}-m_{\rm F814W}}$ provide
estimates for the distance modulus to B379 ($(m-M)_{0}$) and the reddening value ($E(B-V)$):
$\delta m_{\rm F814W}=(m-M)_{0}+A_{\rm F814W}$, and $\delta c_{m_{\rm F606W}-m_{\rm F814W}}=A_{\rm
F606W}-A_{\rm F814W}$. The reddening law from \citet{car89} is employed in this paper. The
effective wavelengths of the ACS F606W and F814W filters are $\lambda_{\rm eff}=5918$ and 8060
\AA~\citep{sirianni05}, so that from \citet{car89}, $A_{\rm {F606W}}\simeq 2.8\times E(B-V)$ and
$A_{\rm {F814W}}\simeq 1.8\times E(B-V)$ \citep[see][for details]{bm07}. The reddening value and
distance modulus for B379 obtained in this paper are: $E(B-V)=0.08$ and $(m-M)_{0}=24.44\pm0.10$,
where the uncertainty is the standard error of the mean.

The best-fitting Padova isochrone can be seen in Figure 1: with the metal abundance $0.009$ in $Z$
(or $-0.325$ in $\rm [M/H]$) and 11.0 Gyr in age. The age of B379 obtained in this paper is
$11.0\pm1.5$~Gyr, where the uncertainty is the standard error of the mean.

\begin{figure*}
\figurenum{1}\hspace{2.0cm}\epsscale{1.0} \rotatebox{-90}{\plotone{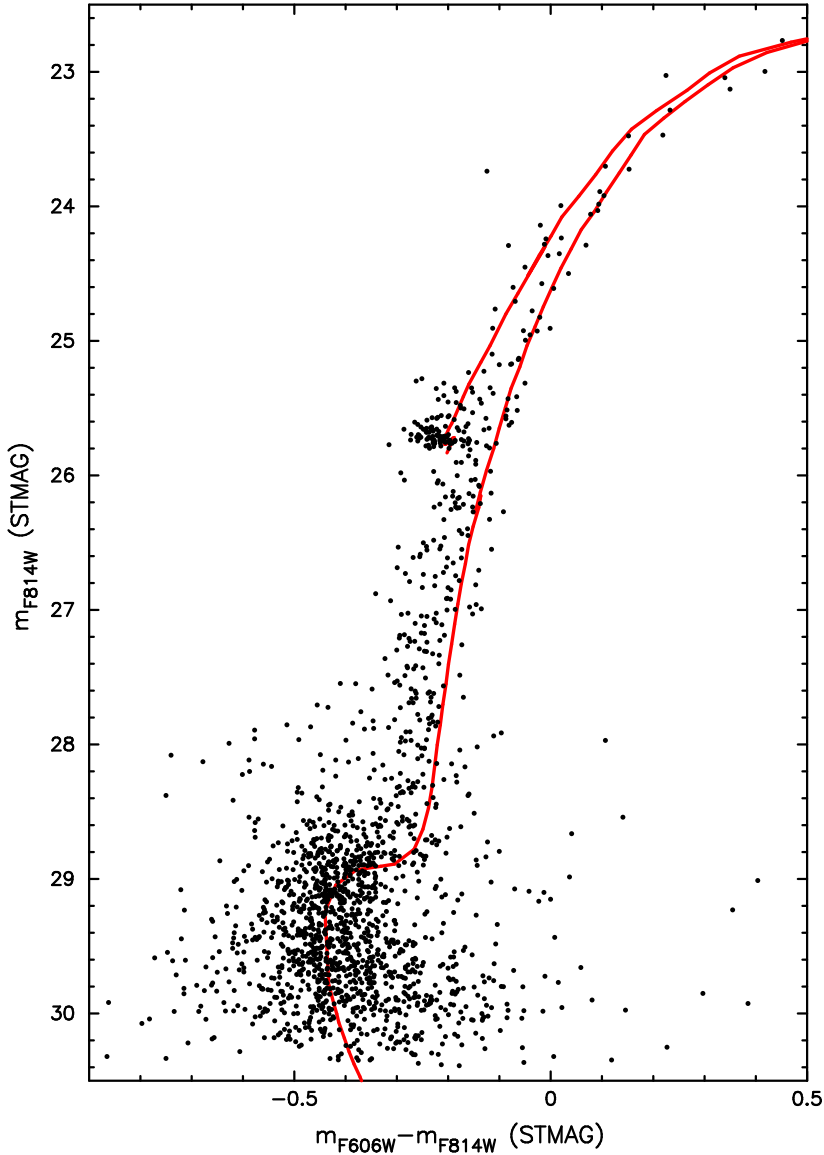}} \vspace{0cm}
\caption{Best-fitting Padova isochrone overplotted on the cluster CMD. The isochrone has the metal
abundance $\rm [M/H]=-0.325$ and age 11.0 Gyr. The isocrone has been shifted by $E(B-V)=0.08$ and
$(m-M)_0=24.44$.} \label{fig:onea}
\end{figure*}

The primary purpose of this paper is to obtain the age of B379 by comparing its CMD with
isochrones of the Padova stellar evolutionary models, and to check whether the age of B379
obtained in this paper, is in agreement with the determination of \citet{brown04a}. From
high-resolution stellar spectroscopy it is presented that GCs in both the halo and the bulge of
our Galaxy are $\rm{\alpha/Fe}$ enhanced with the typical values $\rm{[\alpha/Fe]\approx
0.3\pm0.1}$ dex \citep[see][and references therein]{Thomas03}. For M31 GCs, the estimates of
$\rm{[\alpha/Fe]}$ ratios by \citet{b05} and \citet{ppb05} showed it may on average be $\sim
0.1-0.2$ dex lower than in the Milky Way \citep[see also][]{Colucci09}. The Padova stellar
evolutionary models do not provide isochrones with $\rm{[\alpha/Fe]>0.0}$, however, the
luminosities of turn-off, SGB, and of the tip of the RGB are nearly unchanged by varying $\alpha$
enhancement except in the intermediate-age regime, where $\alpha$-enhanced isochrones are slightly
fainter than scaled solar ones \citep[see][and references therein]{Gallart05}. It is generally
known that the turn-off, SGB and lower RGB are the most age-sensitive features of the CMD, so, the
age of B379 obtained based on the isochrones with $\rm{[\alpha/Fe]=0.0}$ will not change when
using the isochrones with $\rm{[\alpha/Fe]>0.0}$.

\subsection{Comparison with the previously published results}

The age of B379 ($11.0\pm1.5$~Gyr) obtained in this paper is consistent with the determination
($10_{-1}^{+2.5}$~Gyr) of \citet{brown04a}. \citet{brown04a} determined the age of B379 by
comparing the observed CMD with isochrones of \citet{vandenberg06}. The result of this paper
confirmed the conclusion of \citet{brown04a} that B379 is 2--3 Gyr younger than the oldest
Galactic GCs. The metallicity of B379 obtained in this paper is $\rm [M/H]=-0.325$. Taking into
account an enhancement of the $\alpha$-capture elements by $[\alpha/\rm Fe]=0.3$ \citep{brown04a},
and using the relation between [M/H], [Fe/H], and $[\alpha/\rm Fe]$ from \citet{Salaris93}, we
derived $\rm [Fe/H]=-0.54$, which is in good agreement with the determination of $\rm
[Fe/H]=-0.53$ of \citet{hfr97} based on the shape of the RGB of the deep CMD observed by the {\sl
HST}/WFPC2.

B379 is located in the M31 halo, so the extinction is mainly from the foreground Galactic
reddening in the direction of M31, which was discussed by many authors
\citep[e.g.,][]{vand69,McRa69,fpc80,fusi05}, and nearly similar values were determined such as
$E(B-V)=0.08$ by van den Bergh (1969), 0.11 by \citet{McRa69} and \citet{hodge92}, 0.08 by
\citet{fpc80}. In addition, \citet{bh00} determined the reddening for each individual cluster
using correlations between optical and infrared colors and metallicity, and by defining various
``reddening-free'' parameters using their large database of multi-color photometry. Finally,
\citet{bh00} determined reddenings for 314 clusters, 221 of which are reliable \citep[see][for
details]{bh00}. For B379, \citet[][also P. Barmby, priv. comm.]{bh00} obtained its reddening value
to be $E(B-V)=0.10\pm 0.05$. It is evident that the reddening value of $E(B-V)=0.08$ obtained in
this paper is consistent with these determinations.

Given the importance of M31 as an anchor for the extragalactic distance scale, many studies have
presented distance determinations to M31 using different methods. \citet{pv87}, \citet{holland98}
and \citet{vilardell06} have given a detailed review. Although the stellar populations located in
different positions in M31 have different distance moduli, the dispersion can be neglected since
the distance of M31 is large enough. For example, \citet{rich05} pointed out that, the clusters in
M31 dispersed over a 20 kpc radius would have up to 0.06 mag random distance uncertainty. So, the
distance modulus to B379 obtained in this paper should be consistent with the distance of M31
previously determined within 0.06 mag random distance uncertainty. Now we compared our
determination with the most recent and/or important measurements. \citet{Freedman90} derived the
mean distance modulus to M31 to be $(m-M)_{0}=24.44\pm0.13$ based on the Cepheids in Baade's
fields I, III, and IV \citep{bs63,bs65} observed using the Canada-France-Hawaii Telescope (CFHT).
\citet{holland98} determined the distance moduli to 14 M31 GCs by fitting theoretical isochrones
to the observed RGBs including B379. The distance modulus to B379 obtained by \citet{holland98} is
$(m-M)_{0}=24.45\pm0.07$. \citet{sg98} estimated the distance modulus to M31 as
$(m-M)_{0}=24.471\pm0.035$ by comparing the red clump stars with parallaxes known to better than
10\% in the Hipparcos catalog with the red clump stars in three fields in M31 observed with the
{\sl HST}. A determination of \citet{freedman01} based on Cepheid P--L relation suggests the
distance modulus of $(m-M)_{0}=24.38\pm0.05$ to M31 when they performed the results of the {\sl
HST} Distance Scale Key Project to measure the Hubble constant. \citet{Durrell01} determined the
distance modulus of $(m-M)_{0}=24.47\pm0.12$ to M31 from the luminosity of the RGB tip of over
2000 RGB halo stars in a halo field located about 20 kpc from the M31 nucleus along the southeast
minor axis. \citet{Joshi03} have obtained $R-$ and $I-$band observations of a $13'\times13'$
region in the disk of M31 and derived the Cepheid period--luminosity distance modulus to be
$(m-M)_{0}=24.49\pm0.11$. \citet{brown04b} determined the distance modulus of $(m-M)_{0}=24.5\pm
0.1$ to M31 based on brightness of 55 RR Lyrae stars detected on the {\sl HST}/ACS images of
$\sim$ 84 hr (250 exposures over 41 days). \citet{McConnachie05} derived the distance modulus to
M31 to be $(m-M)_{0}=24.47\pm0.07$ based on the method of the tip of the RGB observed using the
Isaac Newton Telescope Wide Field Camera (INT WFC). \citet{ribas05} derived the distance modulus
of M31 as $(m-M)_{0}=24.44\pm0.12$ from an eclipsing binary. Very recently, \citet{sarajedini09}
presented the {\sl HST} observations taken with the ACS WFC of two fields near M32 located $4-6$
kpc from the center of M31, and identified 752 RR variables with excellent photometric and
temporal completeness. Based on this large sample of M31 RR Lyrae variables, and using a relation
between RR Lyrae luminosity and metallicity along with a reddening value of $E(B-V)=0.08\pm0.03$,
they derived the distance modulus of $(m-M)_0=24.46\pm0.11$ to M31. In order to see clearly, we
list these determinations of M31 distance moduli in Table 1. It is evident that our determination
is in good agreement with the previous determinations.

\section{Summary}

In this paper, we re-determined the age of the M31 GC B379 by fitting its deep photometry
extending below the main-sequence turn-off to the isochrones of the Padova group
\citep{Girardi02,Girardi08,Marigo08}. The age obtained in this paper is consistent with the
determination of \citet{brown04a}, and confirms the conclusion of \citet{brown04a} that, B379 is
2--3 Gyr younger than the oldest Galactic GC. This paper also confirms the consistence of the age
scale of B379 between the Padova group isochrones used in this paper and the 2006 VandenBerg
isochrones used by \citet{brown04a}. So, the results' comparison between \citet{brown04a} and
\citet{Ma07} is meaningful. In addition, the metal abundance, reddening value, and distance
modulus obtained in this paper are consistent with the previous determinations.

\acknowledgments We are indebted to the referee for thoughtful comments and insightful suggestions
that improved this paper significantly. We are thankful to Dr. Brown for providing the {\sl
HST}/ACS WFC data for B379. This work was supported by the Chinese National Natural Science
Foundation grands No. 10873016, 10633020, 10603006, and 10803007, and by National Basic Research
Program of China (973 Program), No. 2007CB815403.

\newpage
\begin{table}
\caption{Distance determinations to M31 as presented in the references for
comparison.}\label{t3.tab}\vspace{5mm}
\begin{tabular}{lcc}
\tableline \tableline
  Method & $(m-M)_0$   &  Reference \\
         & [mag]       &            \\
  \tableline
  Cepheids         & $24.44\pm0.13$   & [1] \\
  Red Giant Branch & $24.45\pm0.07$   & [2] \\
  Red Clump        & $24.471\pm0.035$ & [3] \\
  Cepheids         & $24.38\pm0.05$   & [4] \\
  Red Giant Branch & $24.47\pm0.12$   & [5] \\
  Cepheids         & $24.49\pm0.11$   & [6] \\
  RR Lyrae         & $24.5\pm0.1$     & [7] \\
  Tip of the RGB   & $24.47\pm0.07$   & [8] \\
  Eclipsing binary & $24.44\pm0.12$   & [9] \\
  RR Lyrae         & $24.46\pm0.11$   & [10]\\
  CMD              & $24.44\pm0.10$   & [11] \\
\tableline
\end{tabular}\\

[1]:~\citet{Freedman90}; [2]:~\citet{holland98}; [3]:~\citet{sg98}; [4]:~\citet{freedman01};
[5]:~\citet{Durrell01}; [6]:~\citet{Joshi03}; [7]:~\citet{brown04b}; [8]:~\citet{McConnachie05};
[9]:~\citet{ribas05}; [10]:~\citet{ribas05}; [10]:~\citet{sarajedini09}; [11]:~this paper.
\end{table}

\end{document}